\documentclass[journal]{IEEEtran}
\usepackage{cite}
\usepackage{amsmath}
\usepackage{upgreek}
\usepackage{graphicx}
\usepackage{url}
\usepackage{color}
\usepackage{hyperref}
\newcommand{\etal}{{\it et al}}

\begin{document}

\title{\bf A Simple Improvement for Permanent Magnet Systems for Kibble Balances: More Flat Field at Almost No Cost}

\author{Shisong~Li,~%~\IEEEmembership{Senior Member,~IEEE,}
        Stephan~Schlamminger,~%~\IEEEmembership{Senior Member,~IEEE,}
        Qing~Wang%,~\IEEEmembership{Senior Member,~IEEE}% <-this % stops a space
\thanks{S. Li and Q. Wang are with the Department of Engineering, Durham University, Durham DH13JT, United Kingdom. S. Schlamminger is with the National Institute of Standards and Technology (NIST), Gaithersburg, MD 20899, USA. Email:leeshisong@sina.com; stephan.schlamminger@nist.gov; qing.wang@durham.ac.uk.}
}

\markboth{Manuscript accepted by IEEE Trans. Instrum. Meas.}{}
\maketitle

\begin{abstract}
Permanent magnets together with yokes to concentrate the magnetic flux into a cylindrical air-gap are widely employed in Kibble balances. These experiments require a uniform magnetic flux density along a vertical path, typically a substantial fraction of the length of the air-gap. Fringe fields that are present at both ends of the air-gap limit the region where the flux density does not change more than a certain relative fraction (here: $5\times 10^{-4}$) of the flux density in the center of the magnet system. By simply adding an iron ring with a rectangular cross-section to the inner yoke at each end of the air gap, the effects of the fringe fields can be counteracted, and, hence, the length of the region, where the flux density remains within a given tolerance band is increased. Compared to the alternative, employing a taller magnet, the proposed method yields a magnet system with an extended region of a uniform field without significantly increasing the mass of the magnet system. Potential applications include compact and table-top Kibble balances.
We investigate possible adverse effects on the performance of the magnet system caused by the additional rings: magnetic field strength, coil-current effect, and a dependence of the radial field on the radial position in the field. No substantial disadvantage was found. Instead, the method presented here outperformed previously suggested methods to improve the radial dependence of the radial field, e.g., shorter outer yoke.
In summary, adding rings to the inner yoke improves the uniformity of the field without a detrimental effect to function, cost, and form factor of the magnet system.
\end{abstract}


% \fi
\IEEEpeerreviewmaketitle

\section{Introduction}
\IEEEPARstart{T}{he} International System of Units (SI) entered a new phase with the 26th General Conference on Weights and Measures (CGPM) in 2018. All seven base units are finally defined by fundamental physical constants of nature \cite{cgpm2018}. For mass metrology, the kilogram is determined by the Planck constant, $h$, and accordingly, the primary mass realizations will be linked to $h$ by feasible methods. The Kibble balance \cite{kb76}, previously known as the watt balance, is one possibility for realizing mass at different scales. The key idea of a Kibble balance is to establish a link between two forms of virtual power: electrical and mechanical power. The former can be described by the Planck constant $h$ based on quantum electrical standards \cite{qe}, and the latter depends on the mass to be determined. The specific principle of a Kibble balance can be found in recent review papers, e.g., \cite{stephan16}.

The operation of the Kibble balance can be described by two measurement phases: (1) In the weighing phase, an electromagnetic force is created by a current-carrying coil located in a magnetic field. The electromagnetic force balances the weight of the mass to be determined, i.e. $BlI=mg$, where $B$, $l$, $I$, $m$ and $g$ denote the magnetic flux density at the coil position, the wire length of the coil, the current through the coil, the test mass, and the local gravitational acceleration. (2) In the velocity phase, the coil is moved through the same magnetic field with a velocity $v$, and the induced voltage $\mathcal{E}$ in the coil is measured as $\mathcal{E}=Blv$. Since $Bl=mg/I=\mathcal{E}/v$, the virtual power equation, $\mathcal{E}I=mgv$, can be obtained, and the mass is determined as $m=\mathcal{E}I/(gv)$. In both phases, the magnetic field $B$ plays an important role (although it cancels in the equations at the end), and, hence, a good magnet design is essential for a Kibble balance.

\begin{figure}[tp!]
\centering
\includegraphics[width=0.5\textwidth]{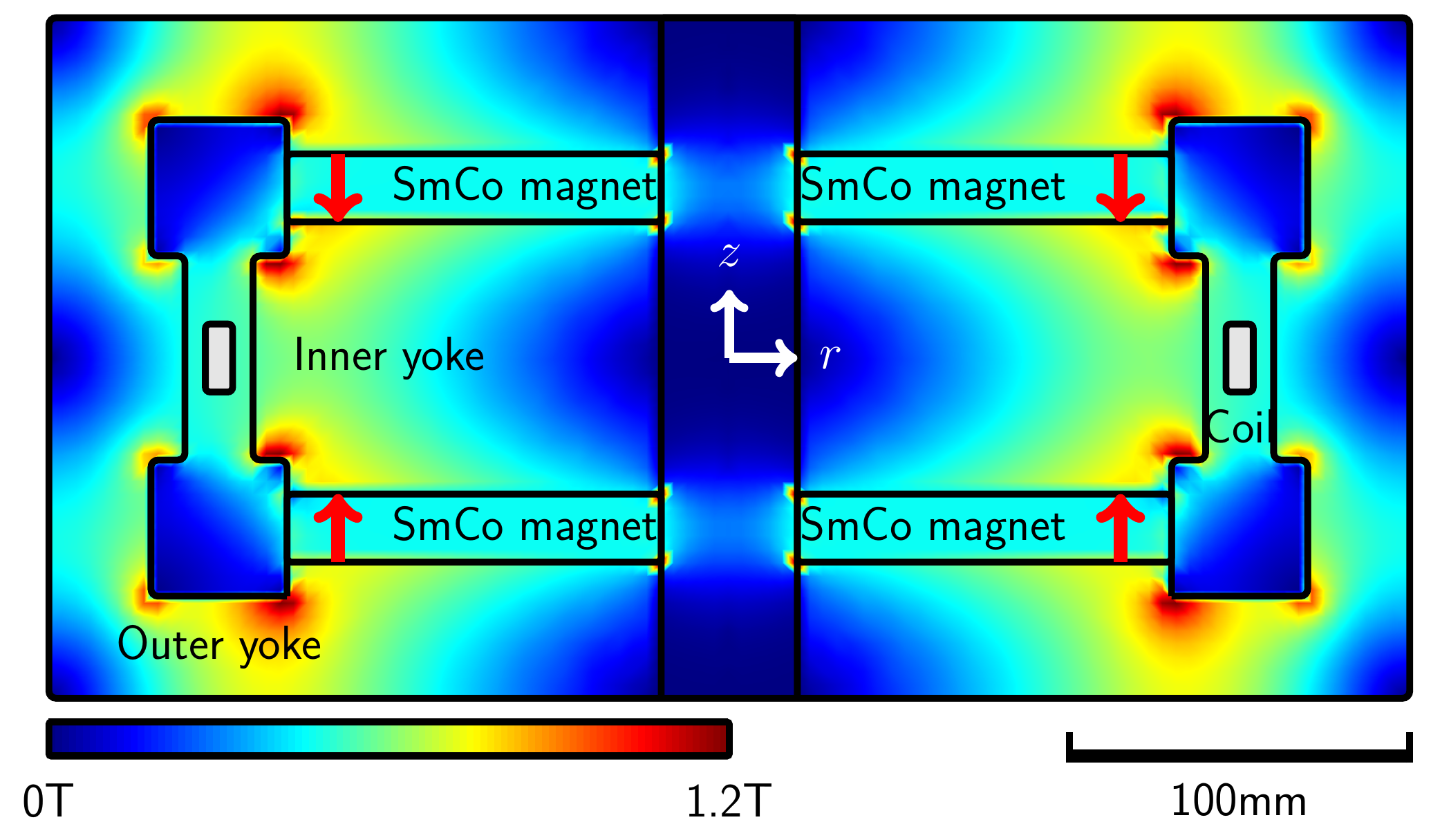}
\caption{The BIPM-type Kibble balance magnetic circuit. The red arrows present the magnetic flux direction of the SmCo magnets. The background color is the amplitude of magnetic flux density. The air-gap width is $\delta=20$\,mm, and the radial magnetic flux density is about 0.5\,T in the center of the air gap. }
\label{fig01}
\end{figure}

At present, all ongoing Kibble balance experiments in the world have chosen permanent magnetic circuits to generate the necessary field required for the measurement \cite{npl,nrc,nist,metas,lne,bipm,nim,msl,kriss,npl2,ume}. A typical symmetrical design, which originated at the  International Bureau of Weights and Measures (BIPM)\cite{bipm}, and has since been chosen widely by researchers at National Metrology Institutes (NMIs)~\cite{nist,nim,kriss,ume}, is shown in Fig.~ \ref{fig01}. The purpose of the magnetic circuit is to compress the magneto-motive force (MMF) supplied by the permanent magnetic material (often SmCo) into a narrow air gap formed by inner and outer yokes and containing the coil. The ideal magnet system provides a strong (fraction of a tesla), vertically uniform, radial magnetic field in the air gap.

The uniformity of the magnetic field along the coil movement trajectory is essential because the induced voltage remains nearly constant along the path when the coil is moved with constant velocity $v$. Clearly, the precise measurement of an almost constant voltage is easier than a widely fluctuating voltage. If the voltage is compensated with a precision voltage source (Zener, Josephson voltage standard), a lower voltmeter range can be chosen. Systematic effects related to the non-linearity of the voltmeter and electrical leakage are reduced. The purpose of the velocity phase is to obtain the value of $Bl$ at the weighing position. It is, usually, obtained by fitting a smooth varying curve to the measurements, i.e. $Bl(z)=\mathcal{E}(z)/v(z)$, taken in the velocity phase.
A longer measurement trajectory reduces the type A uncertainty of the profile fit. Typically, a variation of the magnetic flux density below a few parts in $10^4$ is desired.

For the magnet system discussed here, e.g., the BIPM-type Kibble balance magnet shown in Fig.~\ref{fig01}, the uniform range of the magnetic field is given by the aspect ratio (height/width) of the air gap. Near either end of the air gap, the field uniformity drops quickly due to fringe fields, and the total usable field measurement range is much shorter than the height of the yoke.  Generally, a magnet system with a larger height-to-width ratio exhibits a field that stays uniform over a longer extent. The effect of the fringe field does not reach deep into the gap. Unfortunately, a magnet with a narrow gap has several disadvantages: (1) The clearance around the coil is small, requiring tight control on parasitic (non-vertical) and oscillatory (swing, sway) motions of the coil. (2) The optical elements mounted on the coil have to be small leading to difficulty installing and aligning these components. Small optical elements require a thin laser beam, for which a correction for the Gouy shift must be applied. (3) It is harder to magnetically center a coil in a narrow gap to avoid magnetic torques on the coil. (4) Systematic effects, for example, the coil-current effect are large \cite{ce}. The aspect ratio can also be changed by making the magnet system taller. However, this approach comes with substantial mass and cost for the magnet system. Therefore, an optimization that can suppress the fringe field while keeping the air gap width or magnet size unchanged is desirable.

We propose a small change to the inner yoke to compensate the fringe field and achieve a more extensive range of the uniform field. We show that the proposed method has a minimal negative impact on the main features of the air-gap type magnet system. In section \ref{sec02}, the dependence of the fringe effect, as a function of the air gap width, is discussed. In section \ref{sec03}, a numerical study on the proposed approach is given. Some major concerns of an air-gap type magnet circuit are compared with the original design in section \ref{sec04}. In section \ref{sec05}, experimental measurement results of NIST and BIPM magnetic profiles are compared with that obtained by the finite element analysis (FEA) to ensure that the FEA analysis used is reliable.

\section{Profile flatness dependence on the fringe field}
\label{sec02}
The fringe magnetic field in the air gap can be analyzed similarly to the edge effect of a parallel-plate capacitor. In \cite{fe}, a formula to model the field non-uniformity along the central line in a capacitor is given. According to Appendix A, the formula can be modified to calculate the non uniformity of the radial field in the gap,
\begin{eqnarray}
\mathcal{F}(\delta,d,z)&=&\frac{B_r(z)}{B_r(0)}-1\nonumber\\
&=&-\exp\left[-\left(1+\frac{\pi(d-2|z|)}{\delta}\right)\right]\nonumber\\
&&+\exp\left[-\left(1+\frac{\pi d}{\delta}\right)\right].
\label{eq1}
\end{eqnarray}
Here, $d$ is the height of the air gap, $\delta$ the width of the air gap; $z\in(-d/2, d/2)$ is the coil position in the air gap. For (\ref{eq1}), it was assumed that each yoke has a sharp corner at the end of the gap. An assumption that usually does not hold in reality. However, in the region of interest, the central usable range, where the magnetic profile changes within a few parts in $10^4$, (\ref{eq1}) is a reliable estimator  of the magnetic profile.

We use an FEA to investigate the dependence of the magnetic profile on the dimensions of the air gap. The magnet system analyzed is shown in Fig. ~\ref{fig01}. The external dimensions of the magnet system, radius and height, are both chosen to be 200\,mm. The air gap has a fixed height of $d=60$\,mm and the air gap width is set as a variable. The air gap is always centered around the same coil radius $r_c=150$\, mm. As shown in \cite{nim}, the magnetic flux density in the air gap, $B_r$ is approximately determined by
\begin{equation}
B_r\approx{\mu_0 H_m}/\left({\displaystyle \frac{\delta}{\delta_m}+\frac{S}{S_m}}\right),
\label{eq2}
\end{equation}
where $H_m$ is the coercive field of the permanent magnet ring (in $z$ direction), $\mu_0$ the magnetic permeability of space, $\delta_m$ the height of the permanent magnet ring, and $S$, $S_m$ respectively the half area of the air gap surface ($\approx\pi r_c d$) and the area of the SmCo ring ($S_m=\pi r_{om}^2-\pi r_{im}^2$, with $r_{im}$, $r_{om}$ denoting the inner, outer radii of the SmCo ring, respectively).

Eq. (\ref{eq2}) shows that a smaller air gap width $\delta$ yields a larger magnetic field in the air gap following a nonlinear relation. As shown in the caption of Fig.~\ref{fig01}, the magnetic flux density in the air gap center is about 0.5\,T when the air gap width is set to $\delta=20$\,mm.
Using a material with high relative magnetic permeability ($>1000$) for the yoke, for example, Fe-Ni alloys  \cite{bipmmag}, yields a negligible dependence of the profile on the yoke permeability. For the calculations here, the relative permeability of the yoke has been set to $\mu_r=10\,000$. Note, the magnet designer has to make sure the yoke material is not in saturation to ensure a large value for $\mu_r$.

\begin{figure}[tp!]
\centering
\includegraphics[width=0.5\textwidth]{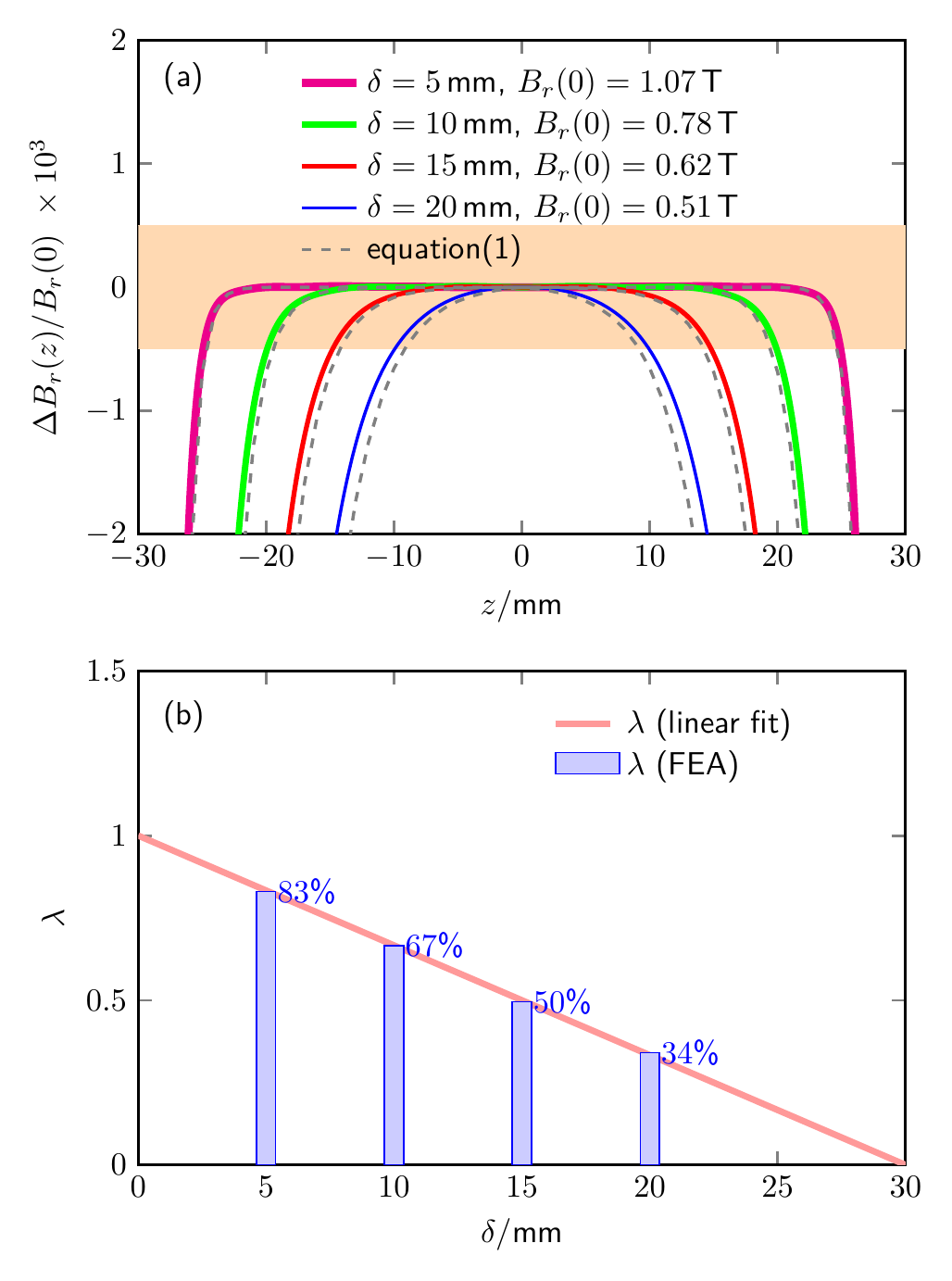}
\caption{The magnetic profile as a function of the air gap width. (a) presents a comparison of the FEA calculation and the analytic model of the relative change of the magnetic flux density with respect to the flux density in the center of the magnet system.  (b) shows the usable range for the velocity measurement when the field change limit is set to $5\times10^{-4}$ (orange region in the upper plot). }
\label{fig02}
\end{figure}

We define the utilization factor as the ratio of the range, with a relative magnetic field change below a limit $\epsilon>0$, over the total height of the air gap $d$, i.e.
\begin{equation}
\lambda=\frac{\displaystyle \Delta z|_{\displaystyle |  \Delta B_r(z)/B_r(0)|\leq\epsilon}}{\displaystyle d},
\label{eq3}
\end{equation}
For the remainder of the article, we set  $\epsilon=5\times10^{-4}$. The utilization factor is the fraction of the air gap in which the magnetic flux density changes relatively by less than 5 parts in $10^4$.

The FEA calculation was performed using four the air gap width $\delta$ the values 5\,mm, 10\,mm, 15\,mm, and 20\,mm. For each case, the radial extent of the air gap spans from  $r_c-\delta/2$ to $r_c+\delta/2$, with $r_c=150$\,mm. The magnetic profile, which is the radial flux density as a function of $z$, for each air gap width is shown in the top panel of Fig.~\ref{fig02} (a). In the same figure, (\ref{eq1}) is plotted. The approximation given in (\ref{eq1}) agrees well (better than one part in $10^4$) with the FEA result. As expected, the agreement between FEA and (\ref{eq1}) is better the larger the aspect-ratio of the air gap, $d/\delta$.

The utilization factors $\lambda$ and the magnetic flux densities in the air gap center $B_r(0)$ for different air gap widths are shown in Fig.~\ref{fig02}(b). The functional dependence of $\lambda(\delta)$ on $\delta$ is linear. We obtain, using a linear regression, $\lambda(\delta)=1-\delta/30\mbox{mm}$. For the largest air gap width, only about 1/3 of the air gap height is usable. The usable range of the profile converges to zero for $\delta=30\,$mm, consistent with the regression equation.

For practical and scientific reasons, some of which were discussed above, a wider air gap is preferred for the considered application. Practically, a wide air gap provides more space to align the coil and to install optical sensors. Tolerances in the machining process will have less effect on the field uniformity for a wide air gap. Scientifically, several systematic effects, e.g., the current effect \cite{ce} of the coil on the result, are smaller for magnets with wider air gaps. A possible solution to increase the trajectory length of the coil sweep in the velocity phase in a uniform field, is to increase the total height of the air gap (and, hence, the magnet). Unfortunately, increasing the height of the air gap will linearly increase the size and exponentially the cost of the magnet. Furthermore, a large magnet is inconvenient for table-top Kibble balances, which need to be transportable. Another solution is to increase the utilization factor without changing the height of the magnet. Increasing the utilization factor leads to a field that is more uniform over a larger trajectory, without necessarily increasing the height (cost and mass) of the magnet. The utilization factor can be increased by preventing some of the flux contributing to the fringe field. The next section describes a very simple way to accomplish this.

\section{A simple modification to the inner yoke}
\label{sec03}
\begin{figure}[tp!]
\centering
\includegraphics[width=0.5\textwidth]{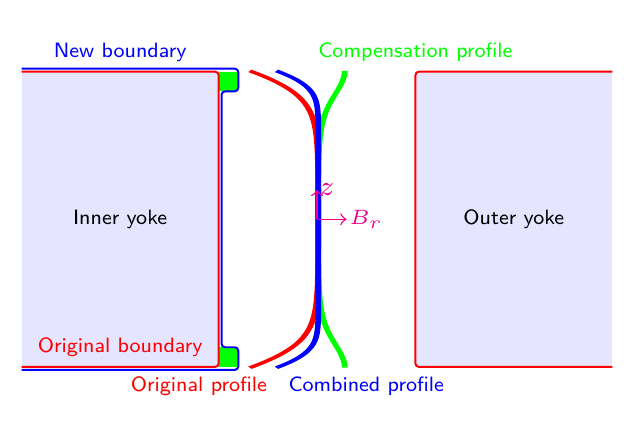}
\caption{Principle of compensation. The inner yoke is designed with two rings at both ends (rectangle in the cross-sectional drawing). The red, green, and blue curves denote the profiles before compensation, the profile created by the rings, and the combined profile, respectively. }
\label{fig03}
\end{figure}

As shown in section~\ref{sec02}, reducing the air-gap width leads to an increase in the magnetic field in the air gap. Since the non-uniformity of the magnetic field is caused by the flux leaking into the fringe field at the two ends of the air gap, a basic idea is to reduce the field gradient by narrowing the air gap at the ends. Changing the magnetic field by modifying the gap geometry is an old idea. For example, in the LNE (Laboratoire national de m\'{e}trology et d'essais) Kibble balance, a linear modification of both inner and outer yokes in the vertical direction has been successfully used to slope an asymmetrical component in the magnetic profile \cite{lnemag}.  In reality, however, a continuously sloped yoke with $\upmu$m level accuracy is difficult to make, and, more importantly, harder to correct once it is over machined \cite{nistmag}. The proposal in this paper is to apply a simple modification of the inner yoke.

The compensation scheme is presented in Fig.~\ref{fig03}. Two rings with a rectangular cross section are added to the inner yoke at the top and bottom of the air gap. Since the inner yoke is most likely machined on a lathe, these steps in the outer radius of the inner yoke can be manufactured easily. The step compensation provides two additional air gaps slightly narrower than the original air gap at both ends of the original air gap. As shown in (\ref{eq2}), a narrower gap allows more flux to go through, and, hence, increases the magnetic field at two ends of the air gap. One way to think about the compensation is illustrated in Fig. \ref{fig03}. The narrower gaps at the ends of the gap produce their own profile, whose slope is opposite to the profile of the original gap. The flat profile occurs as the sum of the compensation profiles and the original profile. 

\begin{figure}[tp!]
\centering
\includegraphics[width=0.48\textwidth]{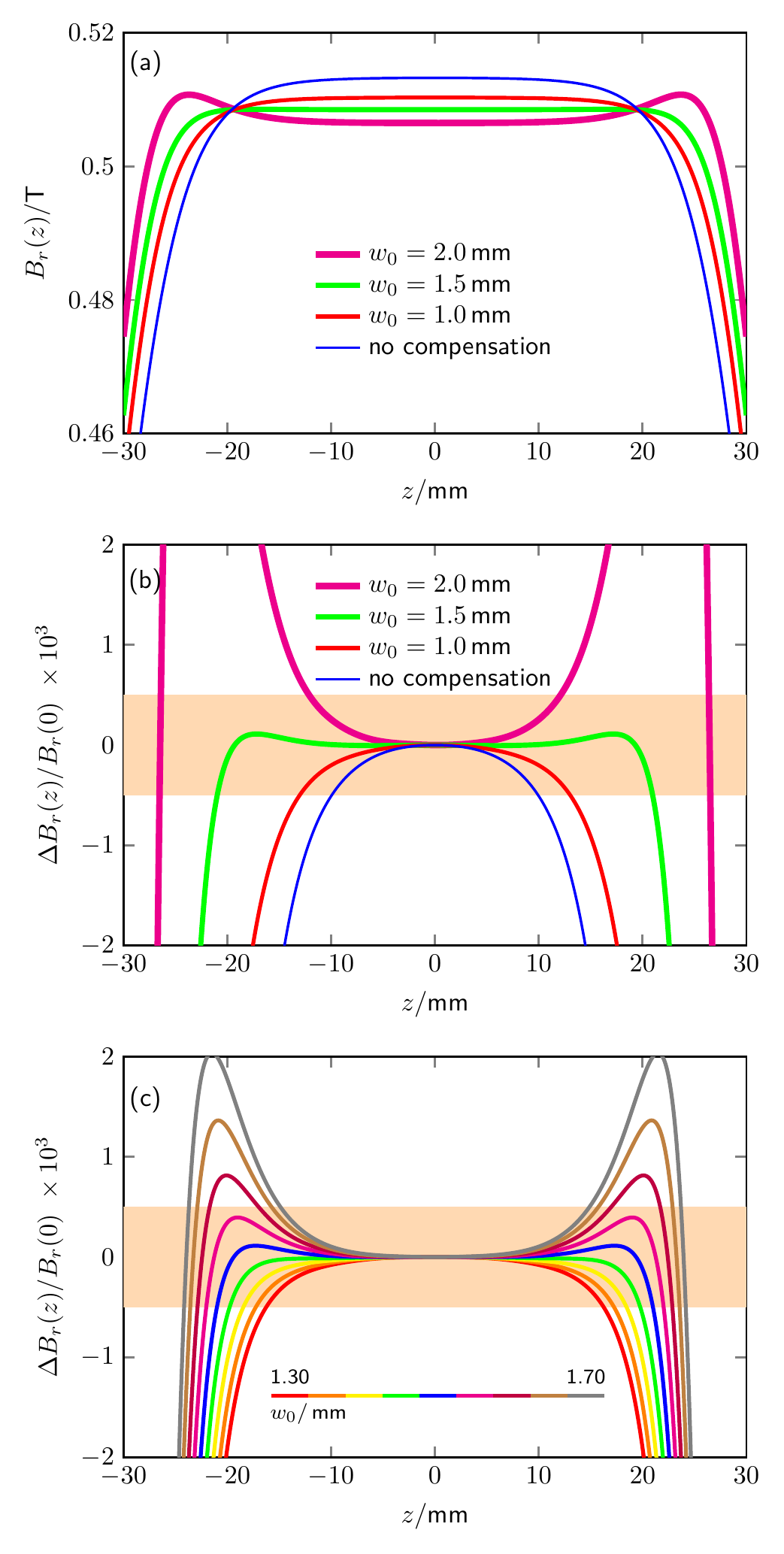}
\caption{Magnetic profiles at different compensation conditions, $h_0=2$\,mm. (a) shows the absolute magnetic field with different step widths. (b) plots the relative field change referred to the magnetic field in the air gap center, $B_r(0)$. (c) is a fine scan of the magnetic profiles with $w_0$ ranged from 1.3\,mm to 1.7\,mm (0.05\,mm step).}
\label{fig04}
\end{figure}

Below, the compensation principle is applied to the magnet system presented in section \ref{sec02} using an air gap width of $\delta=20$\,mm. The heights of both steps are fixed to $h_0=2$\,mm. The FEA analysis is made with three different step widths $w_0$ of 1\,mm, 1.5\,mm, and 2\,mm. The radial component of the flux density in the air gap $B_r(z)$ is calculated for each case, and the result is compared with the original profile, see Fig. \ref{fig04}. As expected, the compensation reshapes the magnetic profile in the air gap by increasing the field at the ends of the air gap and lowering the field in the middle range. The narrower air gaps at the end of the gap provide a smaller reluctance path than the wide air gap. Hence flux from the wide region of the air gap is directed through the narrow regions of the air gap.  Accordingly, the magnetic flux density in the air gap center is slightly reduced.

Of the three different values investigated for $w_0$, the choice  $w_0=1.5$\ produces the profile with the most significant utilization factor, see Fig.\ref{fig04} (b). To obtain the necessary value for $w_0$ that achieves optimal compensation, FEA calculations with $w_0$ ranging from 1.3\,mm to 1.7\,mm in steps of 0.05\,mm were performed. The results are shown in Fig. \ref{fig04}(c). The usable range of the compensated profile can vary slightly at $w_0$ values close to 1.5\,mm. The shapes differ drastically, and experimenters may have different preferences based on other considerations. We prefer a slightly overcompensated case (overcompensation no more than 20\%), achieved with $w_0=1.50$\,mm, i.e., the blue curve in Fig.~\ref{fig04} (c).

The cross-sectional area of the ring in the ideal compensation is $h_0 \times w_0$ =  2\,mm $\times$1.5\,mm. The area is small compared to the cross-sectional area of the original air gap, 60\,mm$\times$20\,mm. At the top of the air gap, its width is 18\,mm  instead of the original 20\,mm. The reduced width will not add significant trouble for placing the coil in the gap or unduly restrict the diameters of laser beams that interrogate the coil.

\begin{figure}[tp!]
\centering
\includegraphics[width=0.5\textwidth]{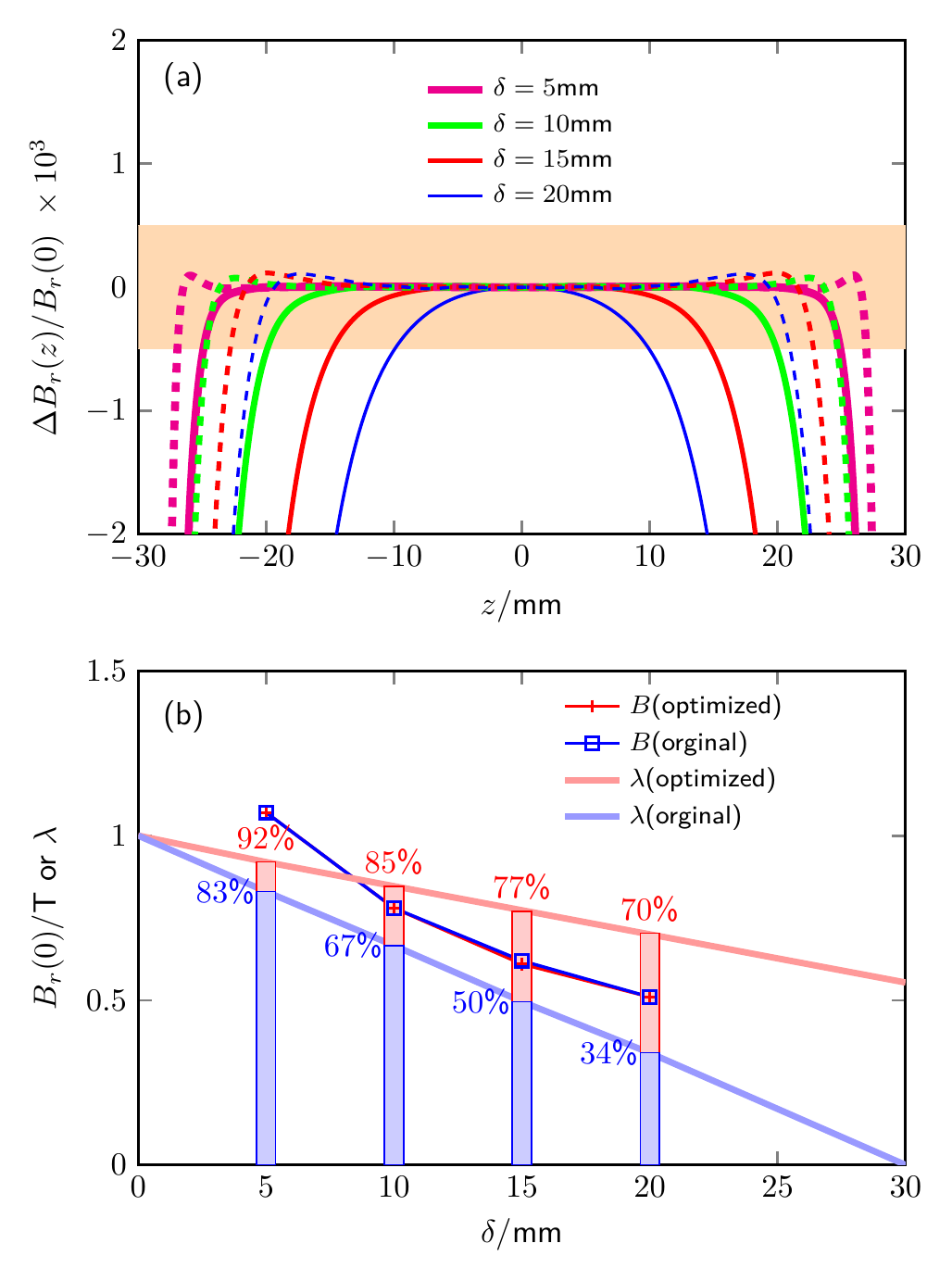}
\caption{Optimal compensation for different air gap widths. (a) presents the comparison of magnetic profiles before and after compensation. The dashed lines are the magnetic profiles after the inner yoke boundary compensation. (b) analyzes the field change in the air gap and usable range improvement after the optimization.}
\label{fig05}
\end{figure}

Another study is made with FEA to check the width of the rings $w_0$  for different air gap widths  $\delta$. The same for values for $\delta$ investigated in section~\ref{sec02} were used here. In each case, the width $w_0$ that produced the largest utilization factor was determined. Fig. \ref{fig05} shows the magnetic profiles without and with ideal compensation for the four different air gap widths, 5\,mm, 10\,mm, 15\,mm, and 20\,mm. The optimized step widths $w_0$ in these four cases are 0.1\,mm, 0.5\,mm, 1\,mm, and 1.5\,mm, respectively. As already seen above, the proposed profile compensation keeps the majority ($>99\%$) of magnetic field strength in the measurement region. By means of regression the utilization range after compensation is found to be $\lambda(\delta)\approx1-\delta/68\mbox{mm}$. The slope has been more than halved compared to the same regression for the uncompensated magnet. For $\delta=30$\,mm, the usable range is still over 50\%. In summary, with the proposed modification of the inner yoke boundary, the usable range of the magnetic circuit has been significantly enlarged. It is now possible to reduce the height of a magnet system without sacrificing the travel range of the coil if a compact magnet system is desired.

Ideally, the magnetic properties and the dimensions of the yoke are precisely known, then the width and height of the ring can be determined with finite element analysis. In that case, the compensation ring and the inner yoke can be machined as one solid piece. A flexible way to realize the compensation discussed here is to manufacture  the compensation rings independent of the inner yoke. Then, the rings can be iteratively adjusted to achieve the optimal of the magnetic profile. The ring could be fixed in place by screws or epoxy after the adjustment. Due to imperfections in the form and the size of each ring, the fit will not be perfect, and after installation, there will be a small gap between the ring and inner yoke. As long as these gaps are much smaller than the main gap of the magnet, they will not affect the compensation produced by the ring, since these parasitic gaps can be seen as in a series magnetic circuit with the main gap reduced by the ring width. To first order, for the circuit analysis, it doesn't matter if the parasitic gap is at the inside or the outside of the ring. 
 For a separated assembly, it is suggested to use materials that reach magnetic saturation at high field. Otherwise, there is the danger that the installation will saturate the material, which can change the magnetic reluctance of the material at lower magnetic field, see~\cite{nistmag}.

\section{Discussion}
\label{sec04}
\subsection{Compensation dependence of step height}
\begin{figure}[tp!]
\centering
\includegraphics[width=0.5\textwidth]{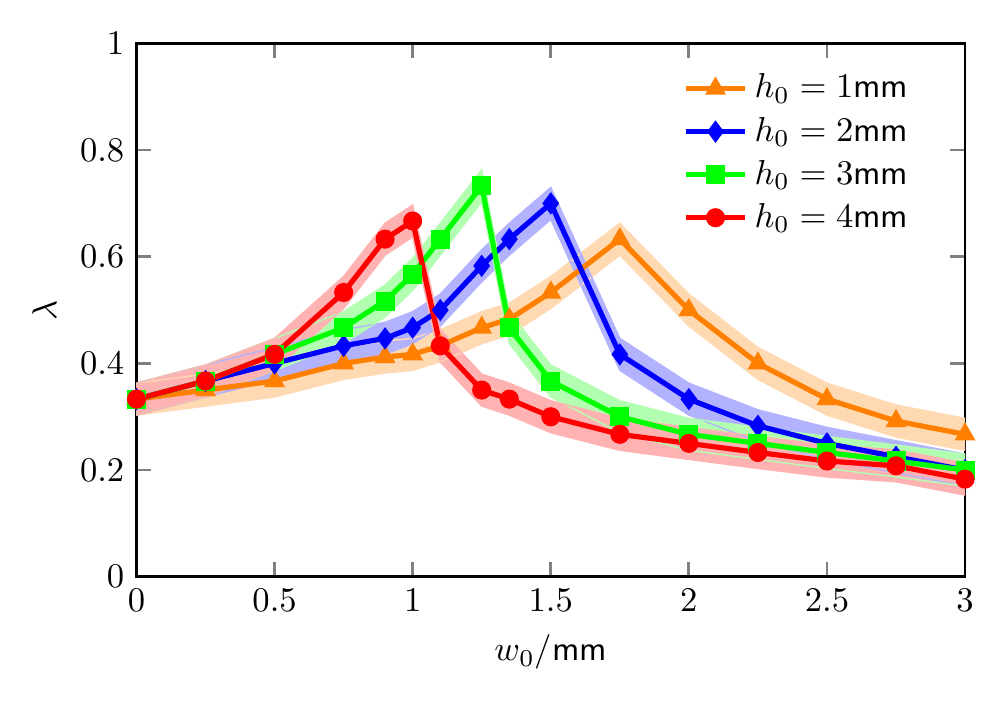}
\caption{The usable magnetic profile range with different compensation step sizes $(h_0,w_0)$. }
\label{fig06}
\end{figure}

The discussion in section \ref{sec03} assumes a fixed height of the compensation step, i.e. $h_0=2$\,mm. The utilization factor as a function of step width $w_0$ is investigated for ring heights $h_0$ of 1\,mm, 2\,mm, 3\,mm, and 4\,mm. The results are shown in Fig. \ref{fig06}. It can be seen that a larger height requires a smaller step width to achieve the best compensation. The usable range for the best compensation becomes less when $h_0$ is chosen to be either too large or too small. In the example discussed here, the largest $\lambda$ can be obtained for values of $h_0$ between  2\,mm and 3\,mm. Two considerations are important for the designer. First, the sensitivity of $\lambda$ on the precise value of $w_0$, i.e., $\partial\lambda/\partial w_0$,  increases with increasing $h_0$. Hence the compensation is more sensitive to tolerances in the machining process. Second, if the narrowing of the gap at the ends of the magnet is a concern, the amount the gap narrows (by $w_0$) can be reduced by choosing a larger $h_0$.

\subsection{Field uniformity along the radial direction}
So far, the dependence of the radial component of the magnetic flux density $B_r$ as a function of $z$ has been discussed. However, the field is a vector (has components $B_r$ and $B_z$) and the coil can move horizontally as well as vertically. One insight that was gained early on in the Kibble balance history is that if the radial component is proportional to $1/r$ or $rB_r= \mbox{constant}$ the result is independent of dimensional changes of the coil caused by resistive heating during weighing. Later~\cite{motion} it was found that such a field is also less sensitive to misalignment of the coil.

According to Maxwell, the divergence of the magnetic field is zero. Expressing $\nabla\cdot \mathbf{B}=0$ in cylindrical coordinates yields
\begin{equation}
    \frac{B_r}{r}+\frac{\partial B_r}{\partial r} +\frac{\partial B_z}{\partial z}=0
\end{equation}
assuming that there is no tangential component $B_\phi=0$ as is the case for a magnet that exhibits cylindrical symmetry. For a field $B_r(r,z)=C(z)/r$, the above equation simplifies to $\partial B_z/\partial z=0$. Hence, a $1/r$ field corresponds to a magnetic field with a constant $z$ component (ideally zero).

It seems that two concentric yokes (inner and outer) will always produce a $1/r$ field. In reality, however,  fringe fields and asymmetries of the boundaries between the inner and outer and the air gap can produce significant vertical components of the magnetic field component, especially when the inner and outer yokes have the same height. In \cite{loweryoke}, two possible modifications to the magnetic circuit are proposed: (1) add permanent magnets and  (2) reduce the height of the outer yoke. Each measure significantly reduces the vertical component of the magnetic field, and, hence, improves the radial dependence of the field. The effect on the utilization factor that these two measures might have is not discussed in \cite{loweryoke}. Here, we investigate the $1/r$ uniformity of $B_r$ with the proposed profile compensation. In addition, the effect of reducing the outer yoke on the utilization factor is studied.

\begin{figure}[tp!]
\centering
\includegraphics[width=0.5\textwidth]{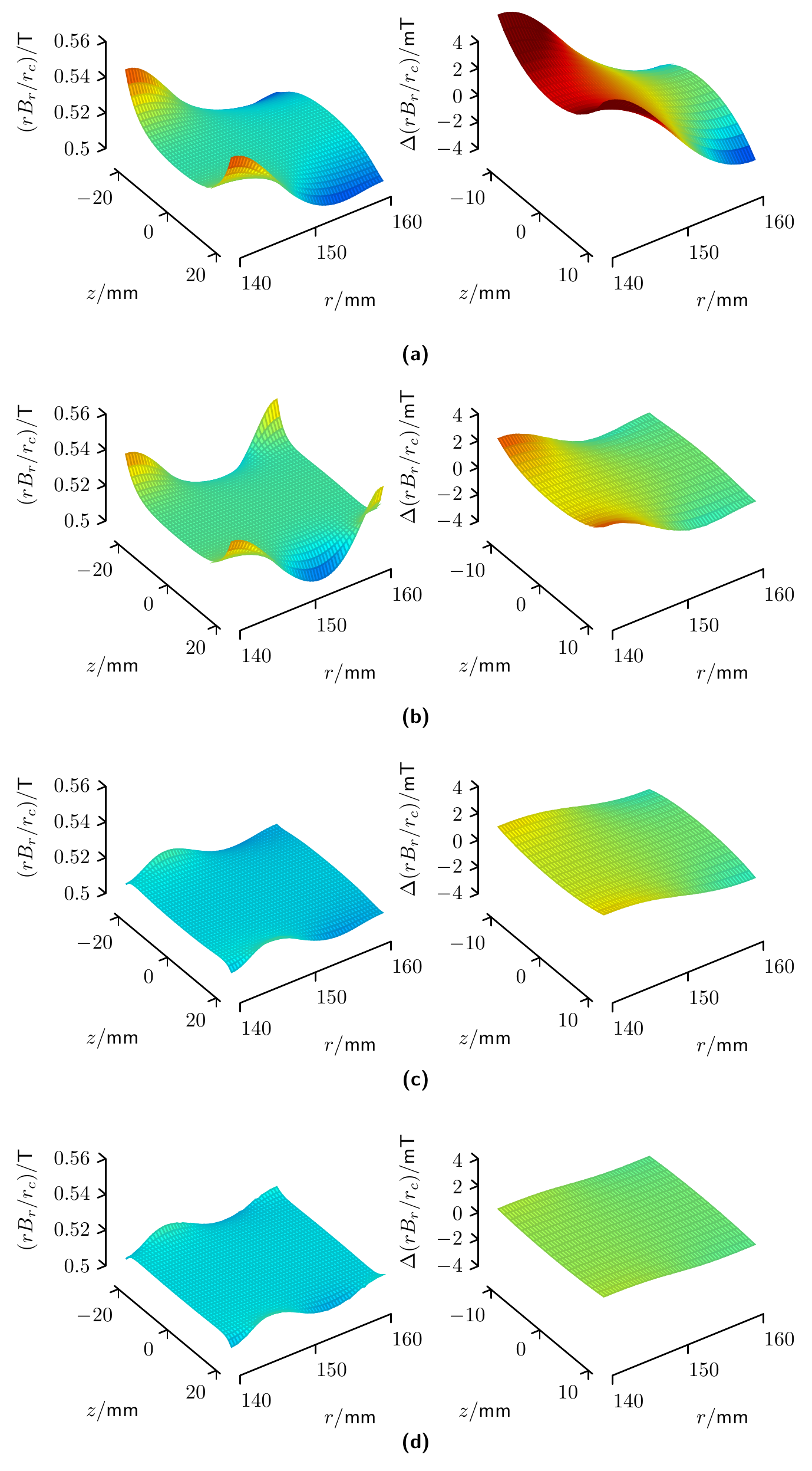}
\caption{The uniformity of the radial magnetic field in the air gap region for four different magnet designs: (a) simple gap; inner and outer yoke have the  same height, (b) height of the outer yoke reduced by 8\,mm, (c)  compensation rings at inner yoke ($h_0=2$\,mm, $w_0=1.5$\,mm), and (d) a combination of (b) and  (c) rings with $h_0=2$\,mm, $w_0=1.5$\,mm  and height reduction of the  outer yoke by 2\,mm. In each row, the right plot is a zoomed version of the left in the central region of the left plot. }
\label{fig07}
\end{figure}

The 2-dimensional flatness of the radial magnetic field is calculated for four configurations: (a) The inner and outer yokes have the same height without compensating rings; (b) The height of outer yoke is 8\,mm shorter as suggested in ~\cite{loweryoke} to reduce $B_z$ at the air gap center; No inner yoke compensation is applied; (c) The inner and outer yokes have the same total height with compensation rings ($h_0=2$\,mm, $w_0=1.5$\,mm) according to section~\ref{sec03}; (d) Combination of conditions (b) and (c); A compensation ring with $h_0=2$\,mm and $w_0=1.5$\,mm) plus the outer yoke is 2\,mm shorter. The desired qualities of the profile in $r$ and $z$ can be visualized by plotting $rB_r(r,z)$. A field that is uniform in $z$ but proportional to $1/r$ in $r$ requires $rB_r(r,z) =\mbox{constant}$; A horizontal plane in a 3d-plot. The units of $rB_r$ are unusual, and hence the numerical values do not provide insight, but the tesla can be regained by dividing the expression by the coil radius $r_c$. Fig.~\ref{fig07} shows plots of $r/r_c B(r,z)$ for the configurations (a)-(d). Part (c) of Fig.~\ref{fig07} impressively shows that profile compensation proposed here, the rings,  is also an efficient way to reduce the vertical magnetic component. It is even more effective in suppressing the vertical component of the magnetic field than reducing the outer yoke height, see part (b) of Fig.~\ref{fig07}. A combination of both methods, see part (d) Fig.~\ref{fig07} achieves the flattest field. Note, when combining both methods with the measures, the height reduction for the outer yoke is much less that than it is only applied alone. In combination, the height of the yoke is only reduced by 2\,mm instead of 8\,mm for configuration (b).

\begin{figure}[tp!]
\centering
\includegraphics[width=0.5\textwidth]{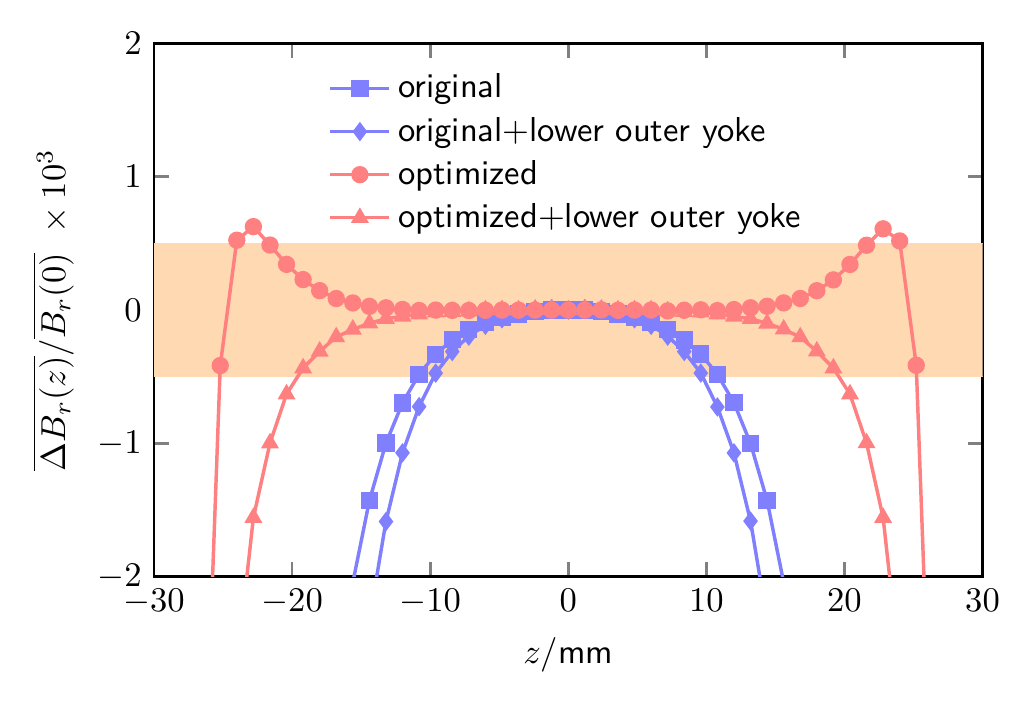}
\caption{The actual magnetic profile determined by the coil. The cross-sectional area of the coil is 10\,mm$\times$10\,mm, centered at $r_c=150$\,mm.}
\label{fig08}
\end{figure}

Knowing the magnetic flux density distribution in the complete air gap allows us to calculate the measured profile in a Kibble balance by integrating over the coil volume,
\begin{equation}
\overline{B_r(z)} =\frac{2\pi}{V} \int\int B_r(r,z)\, r \mbox{d}r\,\mbox{d}z.
\end{equation}
For a coil with rectangular cross-section (width $W$, height $H$), the radial integral is performed from $r_c-W/2$ to $r_c+W/2$, and the vertical integral from $z-H/2$ to $z+H/2$. In this case, the volume of the coil is $V=2\pi r_c H W r_c$. Here, we use $W=H=10$\,mm to obtain the profiles shown in Fig. \ref{fig08}. Two conclusions can be obtained from the result: First, the improvement of the usable magnetic profile range by inner yoke modification is not significantly affected by averaging over the coil volume, and the total flat field range is comparable to that of the central profile at $r_c$. Second, lowering the outer yoke does decrease the uniformity range of the magnetic profile. This can be understood because the equivalent height of the air gap is slightly reduced. Hence, the aspect ratio of the gap is smaller leading to, as shown in section~\ref{sec02}, a less uniform field.

\subsection{Change of the coil-current effect}
The current in the coil during weighing mode slopes the magnetic profile. This is caused by the change in the coil inductance energy change at different vertical positions~\cite{ce}. This effect yields a bias in force mode when the vertical coil position is different for the two measurements in the weighing phase, mass-on and mass-off. The size of the bias is proportional to the slope $k$ in the magnetic profile slope produced by the coil current. The slope depends on the air gap size $(\delta,d)$ \cite{ce}. Since the suggested modification of the inner yoke alters the air gap (at least at the top and bottom of the air gap), it is prudent to check the magnitude of the coil current effect.

\begin{figure}[tp!]
\centering
\includegraphics[width=0.5\textwidth]{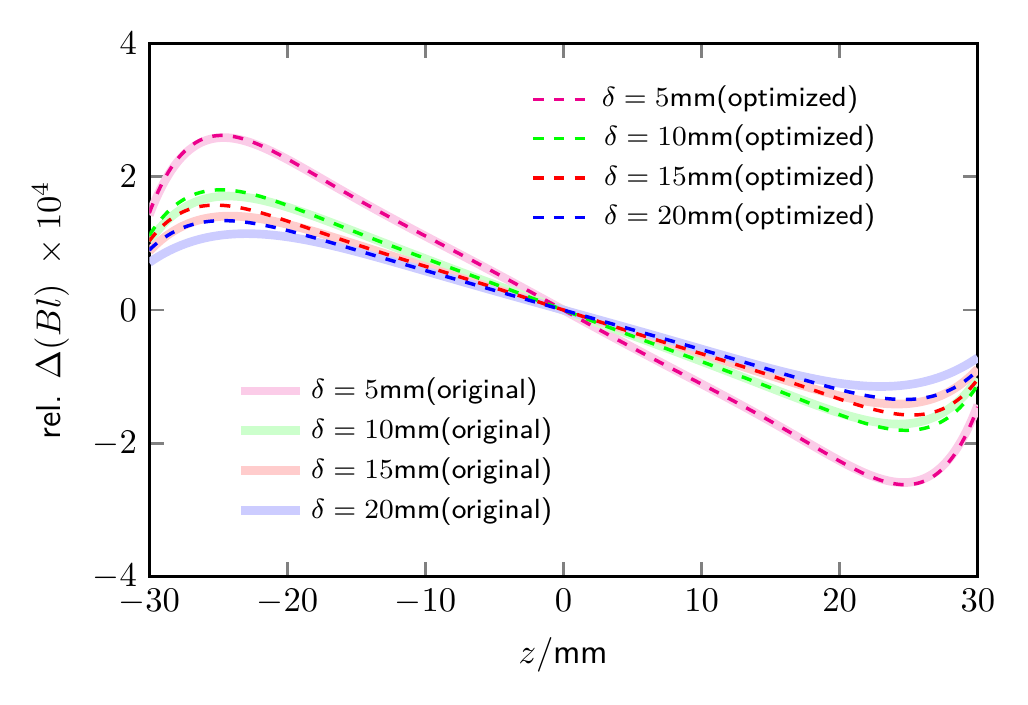}
\caption{Comparison of the weighing magnetic profile change before and after inner yoke boundary compensation. The curves are obtained by calculating the inductance force using parameters shown in Table \ref{tab01}. In the plot only the profiles with positive currents are shown.}
\label{fig09}
\end{figure}

Fig. \ref{fig09} shows the change in the magnetic field profiles as a function of the coil position due to the coil current in the weighing phase. The curves in the figure are obtained by calculating the inductance force along the vertical trajectory.  For the calculation, the total electromagnetic force is set to 4.9\,N, corresponding to the weight of a 500\,g mass. The calculation parameters and the profile slope $k$ at the weighing position ($z=0$) are shown in Table \ref{tab01}. After adding the rings to the inner yoke, the coil current effects differ by less than 1\% from those in the uncompensated designs. The difference can be attributed to the decrease in the magnetic field caused by the compensation. Interestingly, the change in profile is more linear over larger scales after the compensation is applied. For a Kibble balance operating with a single mode  \cite{bipm,ce}, better linearity of the measurement trajectory simplifies the velocity profile fit in the data processing.

\begin{table}
\centering
\caption{The magnetic profile change due to the coil current.}
\label{tab01}
%\begin{indented}
\begin{tabular}{c|ccc|ccc}
\hline\hline
 &  \multicolumn{3}{c|}{Original} & \multicolumn{3}{c}{Optimized}\\
\hline
$\delta$    & $B_r(0)$  &$NI$   &$k$    & $B_r(0)$  &$NI$   &$k$\\
mm  &   T       &   A       &   $10^{-6}$/mm&   T       &   A       &   $10^{-6}$/mm\\
\hline
5    &    1.072    &    4.848    &    -11.154    &    1.071    &    4.856    &    -11.172    \\
10    &    0.779    &    6.675    &    -7.739    &    0.775    &    6.708    &    -7.776    \\
15    &    0.617    &    8.425    &    -6.537    &    0.612    &    8.489    &    -6.587    \\
20    &    0.513    &    10.130    &    -5.894    &    0.509    &    10.225    &    -5.963    \\
\hline\hline
\end{tabular}
%\end{indented}
\end{table}

\section{Experimental check on FEA calculations}
The major purpose of this paper is to provide a simple solution for a compact and efficient magnet system. Since building a Kibble balance magnet is expensive and time-consuming, it is not our interest to construct a prototype and give an experimental verification of the proposed technique. However, we would like to address the concern about the validity of the FEA calculations. To check that, the theoretical profile calculations of the BIPM and NIST-4 magnet systems are compared to experimental results. The comparison result will measure the accuracy of the FEA calculations.

\begin{figure}[tp!]
    \centering
    \includegraphics[width=0.5\textwidth]{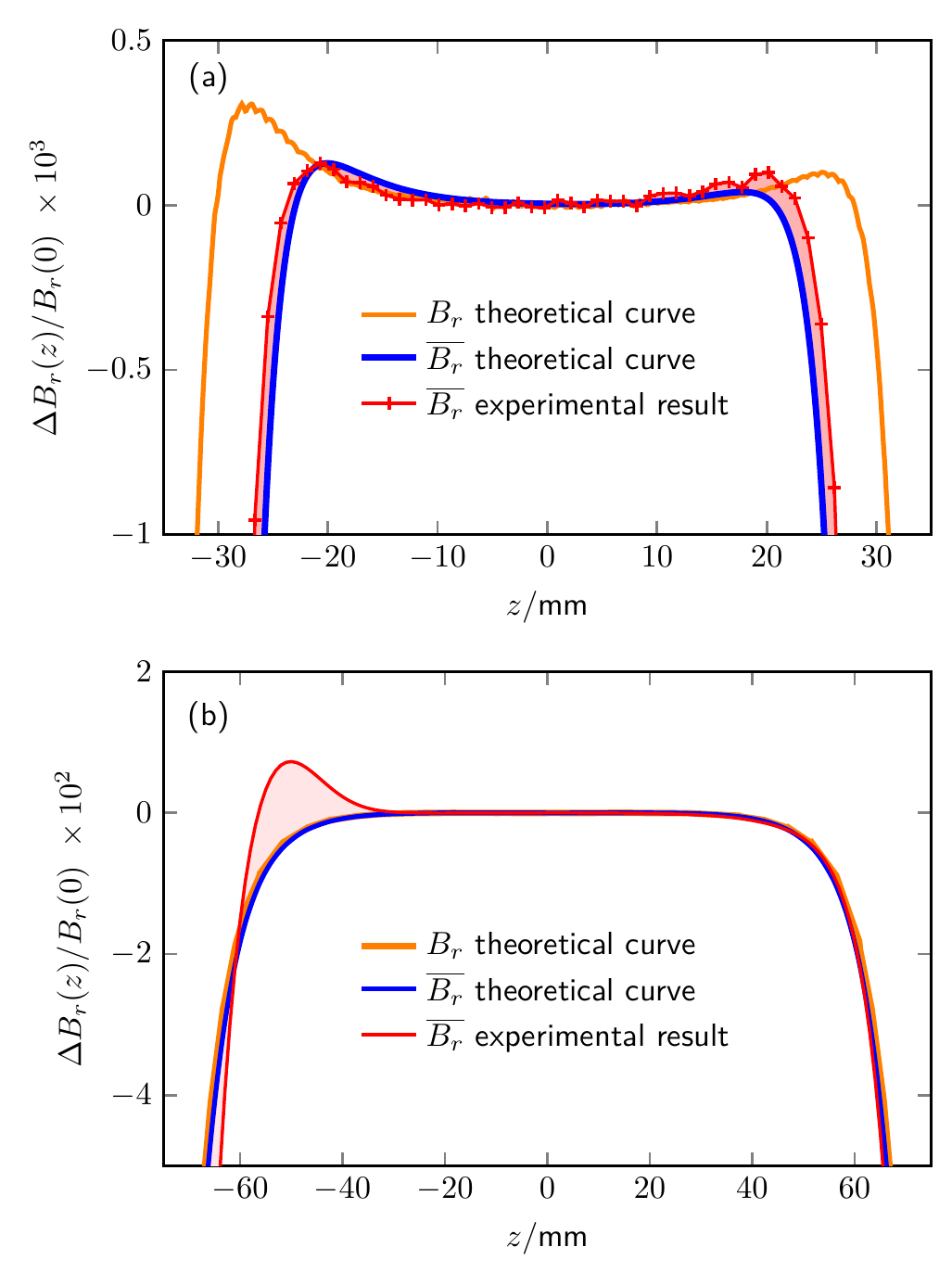}
    %~~\includegraphics[width=0.44\textwidth]{figure11.pdf}
    \caption{Comparison of the profile obtained by simulation and measurement. The results in (a) are performed on the BIPM magnet system \cite{bipmmag}. The theoretical $B_r$ curve is the magnetic flux density distribution along the vertical direction, with a mean radius 126\,mm. The asymmetry is caused due to the missing top cover of the magnetic circuit. $\overline{B_r}$ is the magnetic profile averaged over the coil, for the BIPM system a height of 20\,mm. The experimental result is taken from a voltage-velocity ratio measurement, presented in \cite{bipmmag}. (b) is based on the NIST-4 magnet system \cite{nistmag}. The experimental result was obtained by a gradient coil measurement \cite{nistmag}. The average window for $\overline{B_r}$ is chosen as the height of each gradient coil, i.e. 10\,mm. The mean radius for the simulation equals to the gradient coil radius 217.5\,mm. In both plots, the red shadow denotes the difference between the FEA calculation and the measurement result.}
    \label{fig10}
\end{figure}

The measured and simulated profiles are shown in Fig. \ref{fig10}. The data in Fig.~\ref{fig10}~(a), is obtained from the BIPM magnet system. Details of its construction can be found in \cite{bipmmag}. The air gap is 13\,mm wide and 82\,mm tall (on average). Since the BIPM magnet is not completely closed (the top cover is not yet installed), the profile is not entirely symmetric around the center $z=0$. The asymmetry is observed in the experimental measurement, as well as in the simulation. The quantity $\overline{B_r}$ is obtained by averaging $B_r$ over a region of $z\pm 10\,$mm to match the averaging effect of the coil in the measurement. The result from the simulation agrees well with the measurement. From both curves, the utilization factor $\lambda$ can be calculated. The relative difference in $\lambda$ of the simulation and the experiment is only 2.5\%.  Fig.~\ref{fig10}~(b) shows a similar comparison for the magnet of the NIST-4 system. The detailed design of the NIST-4 magnet system is given in \cite{nistmag}. The air gap 30\,mm in wide and 150\,mm tall. The experimental measurement result was obtained by a pair of 11.5\,mm vertically separated, 10\,mm high gradient coils, therefore, the average window for calculating $\overline{B_r}$, in this case, is chosen as 10\,mm. The asymmetry raised on the left section of the profile during the measurement is due to a mechanical change in the gap width \cite{nistmag}. The FEA result on the right half profile agrees well with the measurement, and the difference in $\lambda$ is about 1.5\%.

In summary, the FEA calculation agrees with the experimental measurement result at the percent level, providing confidence in the proposed compensation idea. 

\section{Conclusion}
\label{sec05}
The Kibble balance is an important instrument for the primary realization of mass in the present international system of units. The air-gap type magnetic circuit has been used widely in Kibble balances to generate a uniform and strong magnetic field. In such magnets, the usable range of the magnetic profile is limited by the ends of the air gap, which allow flux to leak into fringe fields. We propose a simple modification to the inner yoke, that contains more flux in the air gap, and hence improves the uniformity of the field. The modification consists of a pair of symmetrical rings added to the inner yoke at both ends of the air gap. The analysis shows that the proposed compensation can significantly improve the usable range of the magnetic profile, without changing the external dimensions of the magnet. The merits of the proposed inner yoke modification are: (1) Easy to  machine, (2) Minimum impact on the magnetic field strength, (3)  Additional suppression of the vertical magnetic field component in the air gap, which leads to an improvement of the $1/r$ dependence of $B_r$ , (4) Insignificant change to the coil current effect. In summary, the proposed compensation enlarges the usable range in the air gap without significantly changing the mass of the magnet system. It allows the design of more compact magnets for (table-top) Kibble balances.

\section*{Appendix}
In~\cite{fe}, the electric field of a disk capacitor is calculated. The capacitance plates are $2s$ apart, and the $x$ axis is in the center of the capacitor with the plates parallel to the $x$ axis. The original publication uses the letter $d$, to avoid ambiguity we choose $s$ instead. The electric potential difference between the two plates is $2V$. The majority of the field points in the $y$ direction. Eq. (20b) in \cite{fe} gives the relative difference of the electric field in the $y$-direction at the center ($y=0$) from the nominal value of $V/s$ as
\begin{equation}
    \delta_y \approx -\exp{\big(-1-\pi \frac{t}{s}\big)}.
    \label{eq:from17}
\end{equation}
For the capacitor discussed in~\cite{fe}, the direction of the electric field in the gap is (mostly) vertical and the gap is oriented horizontally. For the magnet discussed here,  the magnetic field in the gap is (mostly) horizontal and the gap is oriented vertically.
In (\ref{eq:from17}), $t$ denotes the horizontal distance from the edge of the disk. It needs to be replaced by the vertical distance from the end of the gap $d/2-|z|$. The gap width $2s$ of the capacitor needs to be replaced by the gap width $\delta$ of the magnet. Applying both replacements to~(\ref{eq:from17}) yields
\begin{equation}
    \delta_{Br}(z) \approx -\exp{\big(-1-\pi \frac{ d-2|z|}{\delta}\big)}.
\end{equation}
For the purpose of the article, we investigate the difference in radial field at position $z$ from that at $z=0$. Hence, ~(\ref{eq1}) can be obtained from

\begin{equation}
\mathcal{F}(\delta,d,z)=\delta_{Br}(z) - \delta_{Br}(0).
\end{equation}

\end{document}